\documentclass{article} 	
\usepackage{spconf,amsmath,graphicx,amssymb,booktabs,bm,pifont,multirow}
\usepackage{bbding}
\usepackage{pifont}
\usepackage{wasysym}
\usepackage{textcomp,array,booktabs}
\usepackage[hidelinks]{hyperref}
\newcommand*{\email}[1]{\href{mailto:#1}{\nolinkurl{#1}} }

\newcommand{\splitcell}[1]{\begin{tabular}{@{}c@{}}#1\end{tabular}}
\newcommand{\bsplitcell}[1]{$\left[\splitcell{#1}\right]$}

\makeatletter
\def\endthebibliography{%
\def\@noitemerr{\@latex@warning{Empty `thebibliography' environment}}%
\endlist
}
\makeatother

\title{FDN: Finite Difference Network with Hierarchical Convolutional Features for Text-independent Speaker verification}
\name{Jin Li$^{1,2}$, Nan Yan$^{1,2}$, Lan Wang$^{1,2}$}
\address{$^1$CAS Key Laboratory of Human-Machine Intelligence-Synergy Systems, \\
$^2$Guangdong-Hong Kong-Macao Joint Laboratory of Human-Machine Intelligence-Synergy Systems, \\
Shenzhen Institute of Advanced Technology, Chinese Academy of Sciences, Shenzhen, China \\
\email{{li.jin, nan.yan, lan.wang}@siat.ac.cn}
}

\begin{document}
%
\maketitle
\begin{abstract}
	In recent years, using raw waveforms as input for deep networks has been widely explored for the speaker verification system. For example, RawNet and RawNet2 extracted speaker's feature embeddings from waveforms automatically for recognizing their voice, which can vastly reduce the front-end computation and obtain state-of-the-art performance. However, these models do not consider the speaker's high-level behavioral features, such as intonation, indicating each speaker's universal style, rhythm, \textit{etc}. This paper presents a novel network that can handle the intonation information by computing the finite difference of different speakers' utterance variations. Furthermore,  a hierarchical way is also designed to enhance the intonation property from coarse to fine to improve the system accuracy. The high-level intonation features are then fused with the low-level embedding features. Experimental results on official VoxCeleb1 test data, VoxCeleb1-E, and VoxCeleb-H protocols show our method outperforms and robustness existing state-of-the-art systems. To facilitate further research, code is available at https://github.com/happyjin/FDN
\end{abstract}
\section{Introduction}
\label{sec:intro}

The Speaker verification (SV) field involves determining whether an utterance belongs to a claimed speaker or not. With the development of deep learning technology, an amount of speaker verification studies have replaced the acoustic feature extraction process with deep neural networks (DNNs) and use hand-craft features as input \cite{muckenhirn2018towards, jung2018complete, ravanelli2018speaker}. By contrast, many recent studies explore to use of non-hand-craft features such as raw waveforms \cite{jung2019rawnet, jung2020improved, kim2020segment}.

\begin{figure}[!h]
	\includegraphics[width=8cm]{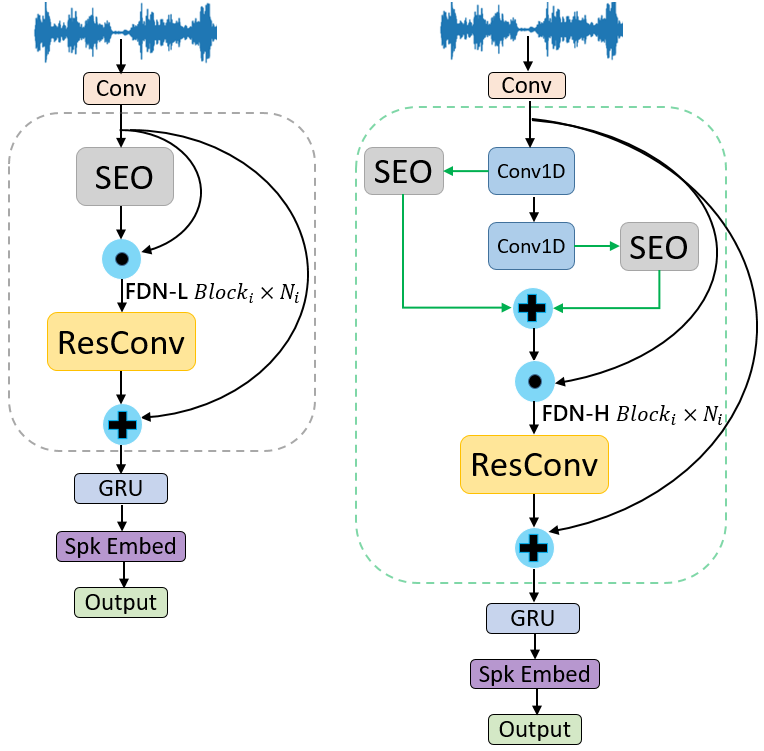}
	\caption{FDN-L and FDN-H building block where Spk Embed represents speaker embedding.}
	\label{fig:seo}
\end{figure}

The Speaker verification (SV) is a critical task in security applications where an utterance belongs to a claimed speaker or not. Recent researches have demonstrated that satisfactory performance can be achieved by end-to-end approaches using raw wave signal\cite{jung2019rawnet, jung2020improved, kim2020segment}. For example, the Sinc-convolution layer of SincNet is successfully applied to the RawNet2 SV system for incorporating the feature scaling to increase the feature diversity \cite{jung2020improved}. Modified RawNet is also used in the short utterance SV task with knowledge distillation from the whole sentence wave signal as a teacher to student \cite{kim2020segment}. However, these systems only consider raw wave signals on the frequency responses without exploring specific speakers' high-level behavior features conveying critical information.

Motivated by psychoacoustics studies \cite{hauser1992fundamental} which show the intonation phenomenon which the rapid final fall at the end of an utterance is a unique property of speaker dispositional feature, language universal regularities,  and inspired by forensic field using utterance-level behavior features as a crucial factor to verify if two audios belong to the same criminal suspect \cite{maher2009audio, machado2019forensic, neustein2012forensic}, we have explored fusion within the model for combining intonation features with speaker's embeddings.  The idea is to represent the rapid final fall phenomenon by calculating the difference of high dimensional global features between the beginning and end of utterance's high dimensional feature, even though the phenomenon was defined by a fundamental frequency decrease in previous studies \cite{gelfer1987simultaneous}. Thus, we propose a novel Finite Difference Network (FDN), which consists of several plug-and-play SEO modules to formulate intonation features by calculating the feature space difference between the beginning and end of utterances using the different kernels. Then, the intonation feature is fused with utterance-level feature to enhance the feature discrimination of different speakers. In addition, a hierarchical finite-difference network (FDN-H or FDN-Heavy) to incorporate features from coarse to fine and improve the robustness of the system (see Figure \ref{fig:seo} FDN-H block) is also explored.


\section{System architecture}
Our proposed speaker verification system consists of a feature extraction module, temporal pooling module, and classification backbends.   

The FDN-L or FDN-H is used for the feature extraction module that extracts features directly from the raw waveform and generates speaker embedding for the current utterance. The underlying hypothesis is that the data-driven approach can learn more discriminative features in the deep neural networks than hand-craft features when a large number of available data can be provided. In addition, the intonation of the speaker can be extracted using the SEO module by computing utterance's variance between the beginning and ending and then incorporated into the utterance features to improve the embedding discrimination and robustness. The details of FDN-L and FDN-H will be described in section \ref{sec:fdn}. Then the GRU layer aggregates frame-level utterance embeddings into an utterance-level feature. A fully connected (FC) is used to output the speaker vectors of the present utterance according to the input of the utterance-level feature.

\begin{figure}[h!]
	\includegraphics[width=8cm]{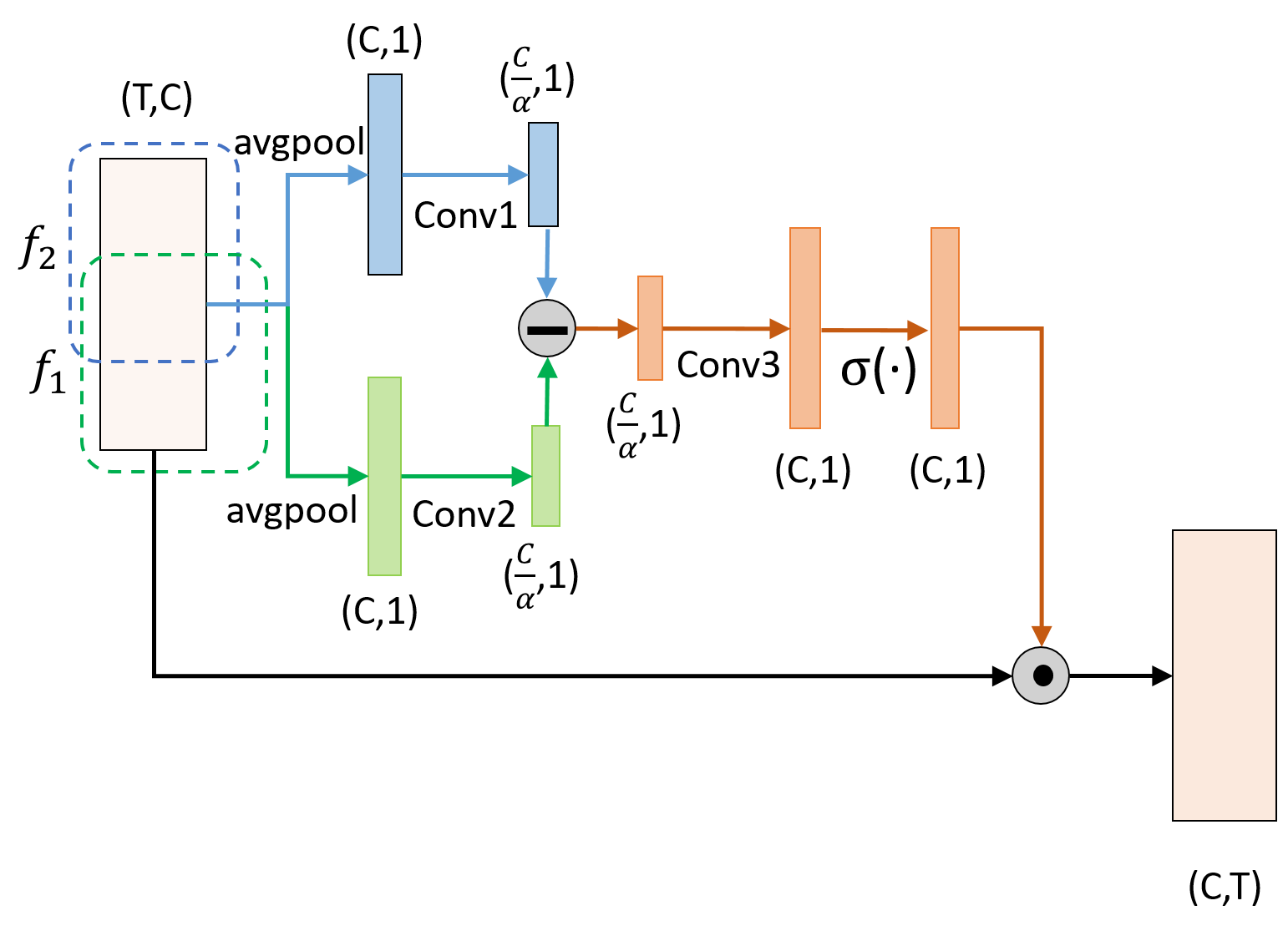}
	\caption{Speed Enhancement Operator pipeline.}
	\label{fig:workflow}
\end{figure}

\begin{table}[h!]
	\caption{Architecture of FDN-L. Batch normalization and LeaklyReLU are applied before the convolution layer in the ResConv, except for the first block \cite{he2016identity}.}
	\label{tab:params}
	\centering
	\setlength\tabcolsep{0.9pt} 
	\setlength{\extrarowheight}{1pt}
	\begin{tabular}{lcc}
		\toprule
		Layer                 & Input: raw waveform (1,T) & \multicolumn{1}{l}{Output size}                \\ \hline
		\multirow{3}{*}{Conv} & Conv(3,3,128)             & \multirow{3}{*}{$(\text{128}, \frac{\text{T}}{\text{3}})$} \\
		& BN                        &                            \\
		& LeaklyReLU                &                            \\ \hline
		FDN-L Block\textsubscript{0}           & \bsplitcell{SEO \\ \bsplitcell{Conv(3,1,128) \\ Conv(3,1,128) \\ MaxPool(3)}ResConv\textsubscript{0}}\texttimes2                       & $ (\text{128}, \frac{\text{T}}{\text{27}})$                  \\ \hline
		FDN-L Block\textsubscript{1}           & \bsplitcell{SEO \\ \bsplitcell{Conv(3,1,256) \\ Conv(3,1,256) \\ MaxPool(3)}ResConv\textsubscript{1}}\texttimes4                       & $(\text{256}, \frac{\text{T}}{\text{2187}} )$               \\ \hline
		GRU                   & GRU(1024)                 & 1024                       \\ \hline
		Sperker               & \multirow{2}{*}{FC(1024)} & \multirow{2}{*}{1024}      \\
		embedding             &                           &                            \\ \hline
		Output                & FC(6112)                  & 6112                       \\ \toprule
	\end{tabular}
\end{table}

\section{Finite difference network}
\label{sec:fdn}
We propose a finite difference network to incorporate the intonation of the speaker in the system. The FDN uses an SEO to split an audio feature into the beginning and ending sub-features along the time axis. And then, the difference between the two sub-features is computed by the subtraction operator after global average pooling. Finally, the intonation of the speaker is incorporated into the input raw waveform by multiplication of weighed channel after sigmoid function with SEO input feature. The SEO operator can be illustrated in Figure \ref{fig:seo}. To further improve the robustness and performance, we also propose an FDN-H that contains a hierarchical architecture (see FDN-H Block in Figure \ref{fig:workflow}).

For the FDN-L module, Let $f\in \mathbb{R}^{C\times T}$ be a feature map of a frame-level representation after Conv layer, where $C$ is the number of channels and $T$ is the sequence length in time. In the SEO, the $f$ is first to split into two adjacent features $f_1\in \mathbb{R}^{C\times T'}$ and $f_2\in \mathbb{R}^{C\times T'}$ by sliding window with window shift. This process can be demonstrated in Figure \ref{fig:seo}. We aggregate two feature maps $f_1$ and $f_2$ along time axis using global average pooling, which produces $f'_1\in \mathbb{R}^{C\times 1}$ and $f'_2\in \mathbb{R}^{C\times 1}$. To reduce the computational complexity and select the most valuable features, two features are fed into two different 1D convolution operators with kernel size 1 in which the number of channels will be compressed. This dimension reduction and the variation of global prosodic calculation can be formulated as:
\begin{equation}
	s = \text{Conv1}(f'_2, W_1) - \text{Conv2}(f'_1, W_2)
\end{equation}
where $W_1$ and $W_2$ are learnable parameters that reduce the number of channels from $C$ to $\frac{C}{\alpha}$. Then the convolution upsamples the number of channels to $C$ and the attention weights are computed by 
\begin{equation}
	s'=\sigma (\text{Conv3}(s_t, W_3))
\end{equation}
where $x'\in \mathbb{R}^{C\times1}$ is attention weights for different channels and $\sigma({\cdot})$ denotes a sigmoid function and $W_3$ is learnable parameter of convolution operator. Finally, the enhanced feature with variation of utterance's beginning and ending is computed by 
\begin{equation}
	u=s'\cdot x
\end{equation}
where $u$ is enhanced feature map. Then, the enhanced feature is input to ResConv (See Figure \ref{tab:params} for details) and a residual connection is linked between input $x$ as well as output of ResConv by addition.

The FDN-H is similar to FDN-L except for the hierarchical architecture. To balance the computational efficiency and performance, we only choose two stacked convolution layer to construct hierarchical architecture. Each convolutional layer has independent SEO (see the FDN-H Block in Figure \ref{fig:seo}). After obtaining attention weights $s'_1$ and $s'_2$ for both two convolutional layers. The final attention weight for hierarchical architecture can be formulated by 
\begin{equation}
	s'_h = \frac{s'_1+s'_2}{2}
\end{equation}
where $s'_h$ is hierarchical attention weight along channel. Then the enhanced utterance-level feature map is computed by 
\begin{equation}
	u_h=s'_h \cdot f
\end{equation}
where $u_h$ is a enhanced hierarchical feature map for intonation of speaker. Then this feature map follows by ResConv and residual connection as same as FDN-L.

\section{Experiments}

\begin{table*}[h!]
	\caption{Results of comparison to state-of-the-art systems on expanded VoxCeleb1-O, VoxCeleb1-E and VoxCeleb-H evaluation protocols.}
	\label{tab:result_compare}
	\centering
	\begin{tabular}{lcccccc}
		\hline
		& Input Feature & Front-end & \multicolumn{1}{l}{Aggregation} & Loss & \multicolumn{1}{l}{Dims} & \multicolumn{1}{l}{EER(\%)} \\ \hline
		\underline{VoxCeleb1-O}     & \multicolumn{1}{l}{}              & \multicolumn{1}{l}{}          & \multicolumn{1}{l}{}            & \multicolumn{1}{l}{}     & \multicolumn{1}{l}{}     & \multicolumn{1}{l}{}        \\ \hline
		Chung \textit{et. al.} \cite{Chung18b}   & Spectrogram                       & ResNet-50                     & TAP                             & Softmax+Contrastive      & 512                      & 4.19                        \\
		Xie \textit{et. al.} \cite{xie2019utterance}    & Spectrogram                       & Thin ResNet-34                & GhostVLAD                       & Softmax                  & 512                      & 3.22                        \\
		Jung \textit{et. al.} \cite{jung2020improved}    & Raw waveform                      & RawNet2                       & GRU                             & Softmax                  & 1024                     & 2.48                        \\ \hline
		{\bf FDN-Light(Ours)} & Raw waveform                      & FDN                           & GRU                             & Softmax                  & 1024                     & \bf 2.44                        \\
		{\bf FDN-Heavy(Ours)} & Raw waveform                      & FDN+Hierarchical                   & GRU                             & Softmax                  & 1024                     & {\bf 2.31}                        \\ \hline
		\underline{VoxCeleb1-E}     & \multicolumn{1}{l}{}              & \multicolumn{1}{l}{}          & \multicolumn{1}{l}{}            & \multicolumn{1}{l}{}     & \multicolumn{1}{l}{}     & \multicolumn{1}{l}{}        \\ \hline
		Chung \textit{et. al.} \cite{Chung18b}   & Spectrogram                       & ResNet-50                     & TAP                             & Softmax+Contrastive      & 512                      & 4.42                        \\
		Xie \textit{et. al.} \cite{xie2019utterance}    & Spectrogram                       & Thin ResNet-34                & GhostVLAD                       & Softmax                  & 512                      & 3.13                        \\
		Jung \textit{et. al.} \cite{jung2020improved}   & Raw waveform                      & RawNet2                       & GRU                             & Softmax                  & 1024                     & 2.57                        \\ \hline
		{\bf FDN-Light(Ours)} & Raw waveform                      & FDN                           & GRU                             & Softmax                  & 1024                     & \bf 2.40                        \\
		{\bf FDN-Heavy(Ours)} & Raw waveform                      & FDN+Hierarchical                   & GRU                             & Softmax                  & 1024                     & {\bf 2.35}                        \\ \hline
		\underline{VoxCeleb1-H}     &                                   &                               &                                 &                          &                          &                             \\ \hline
		Chung \textit{et. al.} \cite{Chung18b}  & Spectrogram                       & ResNet-50                     & TAP                             & Softmax+Contrastive      & 512                      & 7.33                        \\
		Xie \textit{et. al.} \cite{xie2019utterance}    & Spectrogram                       & Thin ResNet-34                & GhostVLAD                       & Softmax                  & 512                      & 5.06                        \\
		Jung \textit{et. al.} \cite{jung2020improved}   & Raw waveform                      & RawNet2                       & GRU                             & Softmax                  & 1024                     & 4.89                        \\ \hline
		{\bf FDN-Light(Ours)} & Raw waveform                      & FDN                           & GRU                             & Softmax                  & 1024                     & \bf 4.55                        \\
		{\bf FDN-Heavy(Ours)} & Raw waveform                      & FDN+Hierarchical                   & GRU                             & Softmax                  & 1024                     & {\bf 4.31}                        \\ \hline
	\end{tabular}
\end{table*}

\subsection{Dataset}
We use the VoxCeleb2 dataset \cite{chung2018voxceleb2} for training and the VoxCeleb1 dataset \cite{nagrani2017voxceleb} to evaluate different protocols. The VoxCeleb2 dataset consists of more than one million utterances from 6112 speakers, and the VoxCeleb1 dataset contains approximately 330 hours of recordings from 1251 speakers for text-independent tasks. 

\subsection{Experimental configurations}
Pre-emphasis \cite{kim2020segment} is employed for raw waveforms before inputting into the network and configured the mini-batch for training by cropping the input utterance to 59049 samples which are around 3.69 seconds. Leaky ReLU activation function \cite{maas2013rectifier} with a negative slope is 0.3. The AMSGrad optimizer \cite{maas2013rectifier} with a learning rate of 0.001 is used and weight decay 1$e^{-4}$ is set. Categorical cross-entropy loss is applied for all output layers. The other parameters related to the system are detailed in Table \ref{tab:params}. The channel reduction ratio of SEO $\alpha$ is set 8. The sliding window size and shift for wave input are 3.19s and 0.5s respectively. There are no data augmentation techniques during the training and test process.

\subsection{Results analysis}

\begin{table}[h!]
	\caption{Number of parameters (million) and VoxCeleb1-O result at different reduction ratio compare with baseline and RawNet2 \cite{jung2020improved}.}
	\label{tab:ratio}
	\centering
	\begin{tabular}{llc}
		\hline
		ratio    & \multicolumn{1}{c}{\#Params} & EER(\%) \\ \hline
		2        & 13.33                   & 2.62    \\
		4        & 13.15                   & 2.45    \\
		\bf	8        & \bf 13.06                   & \bf 2.44    \\
		16       & 13.01                   & 2.46    \\
		32       & 12.99                   & 2.48    \\ \hline
		w/o FDN-L \cite{kim2020segment} & 12.96                   & 3.50    \\
		RawNet2 \cite{jung2020improved}  & 13.38                   & 2.57    \\ \hline
	\end{tabular}
\end{table}
Table \ref{tab:ratio} shows the robust performance which of a range of reduction ratio $\alpha$ under the official VoxCeleb1 evaluation protocol (VoxCeleb1-O). The performance outperforms system without FDN-L. The better results might be since the FDN-L could represent the utterance features and intonation phenomenon of the speaker's dispositional feature that is critical for SV. In addition, channel reduction could select critical filters to reflect intonation in different high dimensional spaces. However, the performance does not increase as the complexity increase as more channels may have redundant information caused to harming the performance. To balance the performance and the computation complexity, The $\alpha=8$ should be selected. Also, our FDN-L is not only lightweight than RawNet2 but also has better performance.

\begin{table}[h!]
	\caption{Ablation study of FDN-L on different blocks.}
	\label{tab:blockN}
	\centering
	\begin{tabular}{ccc}
		\hline
		block 0 & block 1 & EER(\%) \\ \hline
		&         & 3.50   \\
		\checkmark	&         & 2.56   \\
		&  \checkmark       & 2.52   \\
		\checkmark	&  \checkmark       & \bf 2.44   \\ \hline
	\end{tabular}
\end{table}
The ablation study explores the degree to which to insert the FDN-L block and the number of FDN-L blocks inserted into the network (see Table \ref{tab:blockN}). The FDN-L block is attempted to add the block0 and block1 separately and two blocks together. The result shows that the FDN-L preforms increasing results in system. In addition, the FDN-L in both blocks achieves the best performance since more intonation of speaker information is provided.

\begin{figure}[!h]
	\includegraphics[width=8cm]{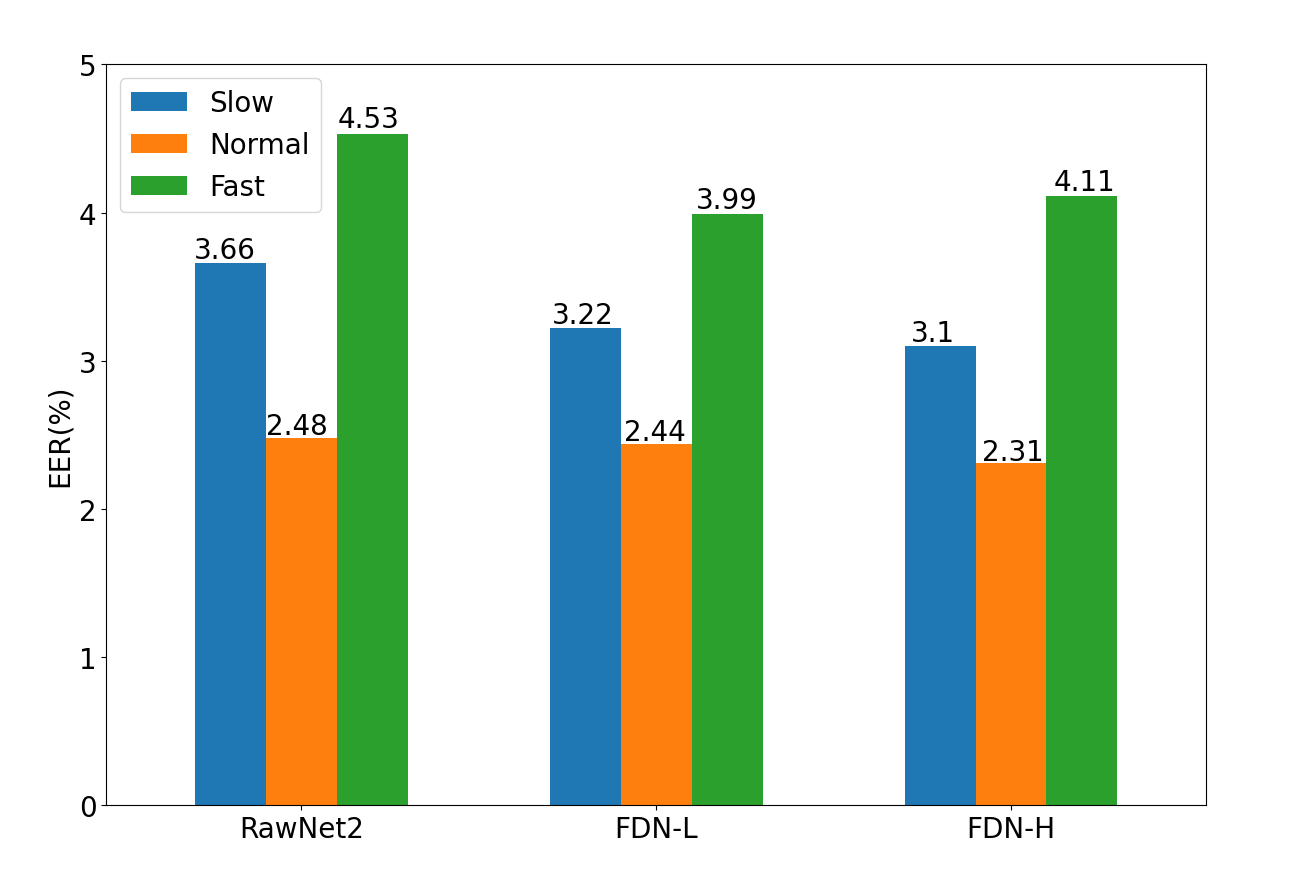}
	\caption{Random speed perturbation on VoxCeleb1-O results.}
	\label{fig:speed_pertub}
\end{figure}
Figure \ref{fig:speed_pertub} shows the result of random speed perturbation on VoxCeleb-O. Slow speed rate ranges from 0.5 to 0.9, fast speed rate ranges from 1.1 to 2.0 and typically keeps the original audios. The results shows our system is more robust than RawNet2 with 3.22 EER for FDN-L under slow speed, 3.99 EER under fast speed, and FDN-H further improves the performance under slow speed. The results demonstrat that the FDN is robust to the speed perturbation when the FDN models the speaker intonation features. However, its performance is a little worse than FDN-L because fast-speed perturbation is hard to extract robust features a hierarchically since less information is contained in the deee layer of CNNsas the receptive field increases.

Finally, Table \ref{tab:result_compare} compares the results obtained in various recent state-of-the-art (SOTA) works using official test protocol VoxCeleb1-O, VoxCeleb1-E, and VoxCeleb1-H. The results show that the proposed FDN-L and FDN-H marginally outperform the SOTA performance, e.g. for FDN-Heavy, EER of 2.31 \% for the official evaluation protocol, and 2.35 \% for the official evaluation of the VoxCeleb-E protocol, 4.31 \% for the VoxCeleb-H protocol.. In addition, the FDN-Heavy outperforms the FDN-Light since hierarchical architecture in FDN-Heavy captures intonation of the speaker from fine to coarse by different convolution layers.

\section{Conclusion}
This paper proposes FDN-L and FDN-H to incorporate the speaker's intonation phenomenon into the system with the input as a raw wave. The FDN is a neural speaker embedding extractor and a plug-and-play module that can be inserted into any network. The experiment results show that our FDN marginally outperforms the current SOTA methods. Future work can be carried out on the duration of intonation.

\section{Acknowledgements}
This work was supported in part by National Key R\&D Program of China (2020YFC2004100), and in part by National Natural Science Foundation of China (NSFC U1736202, NSFC 61771461) and Shenzhen KQTD Project \\ (No. KQTD20200820113106007).

\bibliographystyle{IEEEbib}
\bibliography{strings,refs}

\begin{thebibliography}{10}

\bibitem{muckenhirn2018towards}
Hannah Muckenhirn, Mathew~Magimai Doss, and S{\'e}bastien Marcell,
\newblock ``Towards directly modeling raw speech signal for speaker
  verification using cnns,''
\newblock in {\em 2018 IEEE International Conference on Acoustics, Speech and
  Signal Processing (ICASSP)}. IEEE, 2018, pp. 4884--4888.

\bibitem{jung2018complete}
Jee-Weon Jung, Hee-Soo Heo, Il-Ho Yang, Hye-Jin Shim, and Ha-Jin Yu,
\newblock ``A complete end-to-end speaker verification system using deep neural
  networks: From raw signals to verification result,''
\newblock in {\em 2018 IEEE International Conference on Acoustics, Speech and
  Signal Processing (ICASSP)}. IEEE, 2018, pp. 5349--5353.

\bibitem{ravanelli2018speaker}
Mirco Ravanelli and Yoshua Bengio,
\newblock ``Speaker recognition from raw waveform with sincnet,''
\newblock in {\em 2018 IEEE Spoken Language Technology Workshop (SLT)}. IEEE,
  2018, pp. 1021--1028.

\bibitem{jung2019rawnet}
Jee-weon Jung, Hee-Soo Heo, Ju-ho Kim, Hye-jin Shim, and Ha-Jin Yu,
\newblock ``Rawnet: Advanced end-to-end deep neural network using raw waveforms
  for text-independent speaker verification,''
\newblock {\em arXiv preprint arXiv:1904.08104}, 2019.

\bibitem{jung2020improved}
Jee-weon Jung, Seung-bin Kim, Hye-jin Shim, Ju-ho Kim, and Ha-Jin Yu,
\newblock ``Improved rawnet with feature map scaling for text-independent
  speaker verification using raw waveforms,''
\newblock {\em arXiv preprint arXiv:2004.00526}, 2020.

\bibitem{kim2020segment}
Seung-bin Kim, Jee-weon Jung, Hye-jin Shim, Ju-ho Kim, and Ha-Jin Yu,
\newblock ``Segment aggregation for short utterances speaker verification using
  raw waveforms,''
\newblock {\em arXiv preprint arXiv:2005.03329}, 2020.

\bibitem{hauser1992fundamental}
Marc~D Hauser and Carol~A Fowler,
\newblock ``Fundamental frequency declination is not unique to human speech:
  Evidence from nonhuman primates,''
\newblock {\em The Journal of the Acoustical Society of America}, vol. 91, no.
  1, pp. 363--369, 1992.

\bibitem{maher2009audio}
Robert~C Maher,
\newblock ``Audio forensic examination,''
\newblock {\em IEEE Signal Processing Magazine}, vol. 26, no. 2, pp. 84--94,
  2009.

\bibitem{machado2019forensic}
Thyago~J Machado, Jozue Vieira~Filho, and Mario~A de~Oliveira,
\newblock ``Forensic speaker verification using ordinary least squares,''
\newblock {\em Sensors}, vol. 19, no. 20, pp. 4385, 2019.

\bibitem{neustein2012forensic}
Amy Neustein and Hemant~A Patil,
\newblock {\em Forensic speaker recognition}, vol.~1,
\newblock Springer, 2012.

\bibitem{gelfer1987simultaneous}
Carole~Ellen Gelfer,
\newblock {\em A simultaneous physiological and acoustic study of fundamental
  frequency declination},
\newblock Ph.D. thesis, City University of New York, 1987.

\bibitem{he2016identity}
Kaiming He, Xiangyu Zhang, Shaoqing Ren, and Jian Sun,
\newblock ``Identity mappings in deep residual networks,''
\newblock in {\em European conference on computer vision}. Springer, 2016, pp.
  630--645.

\bibitem{Chung18b}
J.~S. Chung, A.~Nagrani, and A.~Zisserman,
\newblock ``Voxceleb2: Deep speaker recognition,''
\newblock in {\em INTERSPEECH}, 2018.

\bibitem{xie2019utterance}
Weidi Xie, Arsha Nagrani, Joon~Son Chung, and Andrew Zisserman,
\newblock ``Utterance-level aggregation for speaker recognition in the wild,''
\newblock in {\em ICASSP 2019-2019 IEEE International Conference on Acoustics,
  Speech and Signal Processing (ICASSP)}. IEEE, 2019, pp. 5791--5795.

\bibitem{chung2018voxceleb2}
Joon~Son Chung, Arsha Nagrani, and Andrew Zisserman,
\newblock ``Voxceleb2: Deep speaker recognition,''
\newblock {\em arXiv preprint arXiv:1806.05622}, 2018.

\bibitem{nagrani2017voxceleb}
Arsha Nagrani, Joon~Son Chung, and Andrew Zisserman,
\newblock ``Voxceleb: a large-scale speaker identification dataset,''
\newblock {\em arXiv preprint arXiv:1706.08612}, 2017.

\bibitem{maas2013rectifier}
Andrew~L Maas, Awni~Y Hannun, Andrew~Y Ng, et~al.,
\newblock ``Rectifier nonlinearities improve neural network acoustic models,''
\newblock in {\em Proc. icml}. Citeseer, 2013, vol.~30, p.~3.

\end{thebibliography}

\end{document}